\documentclass[final]{aipproc}
\layoutstyle{6x9}
\usepackage{times,colordvi,amsmath,epsfig,float,color,multicol,subfigure,natbib}

\begin{document}

\title{IR and Optical Observations of GRB 030115}

\author{Allyn Dullighan}{
   address={Center for Space Research, Massachusetts Institute of Technology, Cambridge, MA}
}

\author{George Ricker}{
   address={Center for Space Research, Massachusetts Institute of Technology, Cambridge, MA}
}

\author{Nathaniel Butler}{
   address={Center for Space Research, Massachusetts Institute of Technology, Cambridge, MA}
}

\author{Roland Vanderspek}{
   address={Center for Space Research, Massachusetts Institute of Technology, Cambridge, MA}
}


\begin{abstract}

We present an upper limit on the brightness of the afterglow of the long GRB 030115 measured from Infrared (IR) images taken with the Magellan Classic Cam instrument of Ks > 22 at 6.2 days after the burst.  We also present measurments of the host galaxy of GRB 030115 from archival optical and IR HST images taken with the Advanced Camera for Surveys and the Near Infrared Camera and Multi Object Spectrometer 25+ days after the burst.  GRB 030115 is classified as an Optically Dark Burst, as its afterglow was found in the J, H, and K IR bands after a null result was reported in the optical.  This is the first HETE GRB to have its afterglow found initially in the IR.

\end{abstract}

\maketitle

\section{Introduction}

The High Energy Transient Explorer (HETE) detected Gamma Ray Burst (GRB) 030115 (=H2533) \citep{kaw03} on January 15, 2003, at 03:22:34.28 UT.  The Wide Field Monitor (WXM) position was reported to the community over the GRB Coordinate Network (GCN) at 04:33:07 UT, and a 2 arcminute error radius, Soft X-ray Camera (SXC) position was reported at 04:46:34 UT on January 15, 2003.  Followup optical observations were begun early, with observations as deep as the Digital Sky Survey beginning between 1-2 hours after the burst, \citep{att03,cas03,mas03a}, but no conclusive evidence for a fading counterpart was found.  Infrared (IR) observations were begun at $\sim$5 hours after the burst by \citet{lev03} and a fading IR source was reported to the GCN about 15 hours after the burst.  

We made further IR observations of this source with the Magellan 6.5 meter Baade telescope using the Classic Cam instrument on January 23, 2003, but we were not able to analyze the data because of difficulties with the astrometry of the images.  The recent release of archived\footnote{http://archive.stsci.edu/hst/} Hubble Space Telescope (HST) Near Infrared Camera and Multi Object Spectrometer (NICMOS)  and Advanced Camera for Surveys (ACS) observations of GRB 030115 allowed for more accurate astrometric calibration of our images.  Our analysis of the Classic Cam images and the public HST archived data is presented below.  

\begin{figure}
  \includegraphics[width=\textwidth]{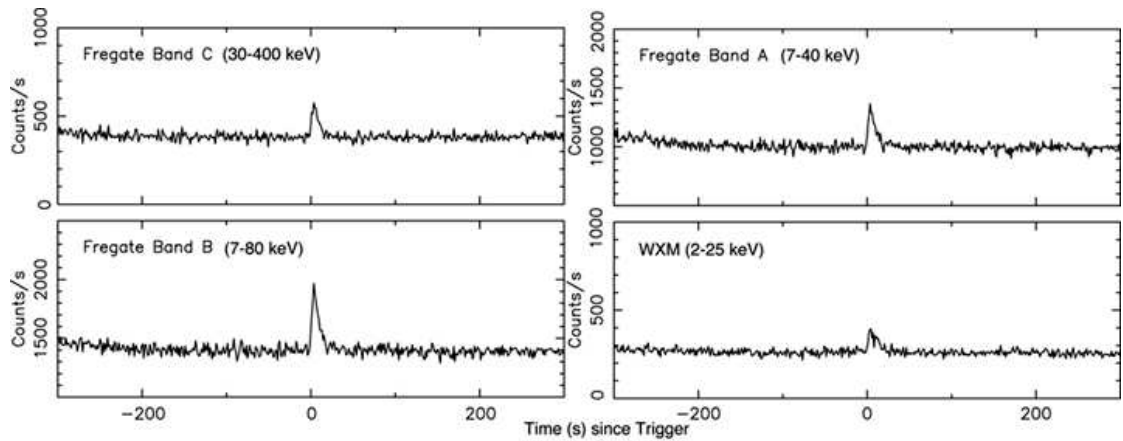}
  \caption{The HETE Fregate and WXM lightcurves of GRB 030115.  The burst lasted $\sim$20sec placing it well within the 'Long/Soft' GRB catagory, and had a waveband evolution typical of these bursts.\label{fig1}}
\end{figure}


\section{Infrared Data and Analysis}

Our IR observations were taken with the Classic Cam IR instrument at the Magellan telescopes at Las Campanas Observatory in Chile.  Classic Cam is a Rockwell NICMOS-3 HgCdTe array with a 256x256 array of 40 $\mu$m pixels.  It was positioned at one of the f/11 foci of the 6.5m Baade telescope, but is no longer in use.  The Classic Cam imager has a maximum field of view of $\sim$30x30 arcseconds, which we used with the K-short filter.  The USNO catalog star $\sim$14'' to the West of the GRB was used to align the telescope with the field.  At the time of our observations, however, engineering issues with the telescope meant that the camera orientation was not well known.  Therefore, with only one bright star in the field of view, it was almost impossible to determine the orientation of the field.  Stacking of  all 45 good images from the data set revealed three other weak sources in the field, but the finder charts available at the time did not clearly resolve other sources in the extremely small $\sim$30x30'' field around the USNO star.

The release of three HST NICMOS F160W filtered images from February 10, 2003, covering the field of GRB 030115 gave an opportunity to orient these images, as the HST images were much deeper, and so clearly resolved the faint sources not seen in the finding charts.  Stacking the three images allowed for a clean removal of cosmic rays in the field in IRAF.  A comparison between the three stacked NICMOS and our Classic Cam image (below) shows the corrected orientation.  Unfortunately, from the image below, you can see that the GRB was below our sensitivity limit for the observation.  A limiting magnitude has been calculated at Ks$\sim$22 for our Stacked Classic Cam image.  IRAF aperture photometry of the stacked NICMOS image gives a magnitude of 24.8 for the host galaxy in the F160W filter ($\sim$H band), when calibrated using the zero point photometry keywords in the HST header files of the images\footnote{see http://www.stsci.edu/hst/nicmos/documents/handbooks/DataHandbookv5/}.

\begin{figure}
  \includegraphics[width=\textwidth]{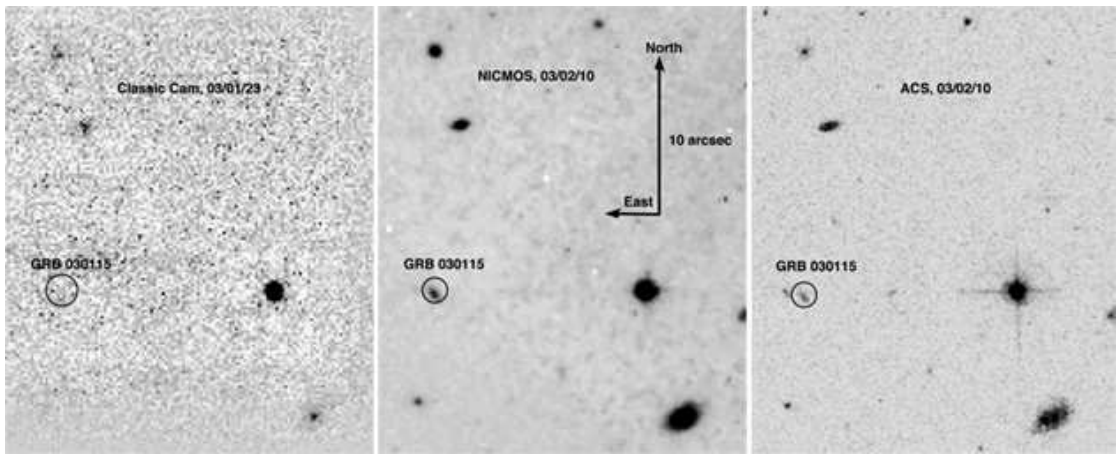}
  \caption{A comparison of the ClassiCAM stacked image (left), the NICMOS stacked image (center), and the ACS two-color image (right) of GRB 030115 at the same astrometric scale.\label{fig2}}
\end{figure}


\section{Optical Data and Analysis}

The IR transient was reported to the GCN at $\sim$16 hours after the burst \citep{lev03}.  At that time earlier observers were able to re-examined their images, and the transient was found in the optical bands at R$\sim$21.5 on January 15, 2003, at 05:25 UT \citep{mas03b} with respect to a nearby USNO-A2 catalog star, or R$\sim$21.9 when recalibrated against the USNO-B1 catalog.  Late time optical observations were made by \citet{gar03} on Jan. 29.4 UT, 2003, and placed the magnitude of the GRB and host at R=25.2, when recalibrated to the USNO-B1 catalog magnitudes.

The public archived ACS images we obtained were taken on February 10, 2003.  There were images available in two filters, F606W and F814W, which are approximately equal to the broad band V and I filters, respectively.  Only one image of good quality was available in each filter, so science quality cosmic ray rejection was not available, but an attempt was made to remove the worst of the cosmic rays by combining the two filters to make the image above.  It clearly shows the small irregular host galaxy of the GRB with no obvious point source remaining, as well as the bright comparison USNO catalog star used for photometric calibration.  The small companion to the host galaxy at 1.3'' to the Northeast, reported by \citet{gar03}, is also resolved in these images.  It is comparatively more blue than the host, only $\sim$0.3 magnitudes dimmer than the host in the F606W filter, and then progressively fainter in the longer wavebands.

IRAF aperture photometry was performed on each of the two ACS images.  A small 5 pixel (0.25'') radius aperture was used to avoid contamination by cosmic rays, which unfortunately led to not all of the galaxy being contained in the aperture.  Our photometry was calibrated against the USNO-B1 catalog star 14 arcseconds to the west of the host galaxy (B=20.640, R=19.450, I=19.690).  The errors in moving between the different filters have not been accurately calibrated, and so the magnitudes reported below should be taken as preliminary.  We find a magnitude of $\sim$26 for the central region of the host galaxy in both filters, F606W and F814W.  This gives a surface brightness of $\sim$24 magnitudes per square arcsecond.


\section{Conclusions}

We can place a new upper limit in the afterglow of GRB 030115 of Ks > 22 on January 23, 2003.  This upper limit is too faint to be consistent with the afterglow decay of $\sim$0.7 between the two earlier Ks measurements reported by \citet{kat03b}, which implies a steepening of the light curve at some time between $\sim$2 and 6 days after the burst.  We also place the magnitude of the afterglow plus host galaxy in F160W ($\sim$H band) at 24.8 at 25 days after the burst.  Optically, the host galaxy of GRB 030115 is seen to be a small, irregular galaxy, perhaps interacting with its companion to the Northeast.  A slight count excess at the position of the GRB in the Digital Sky Survey (DSS) image of the region was reported by \citet{mas03b}, but this is most likely noise, given the faintness of the host galaxy in the ACS images.

\begin{figure}
  \includegraphics[width=\textwidth]{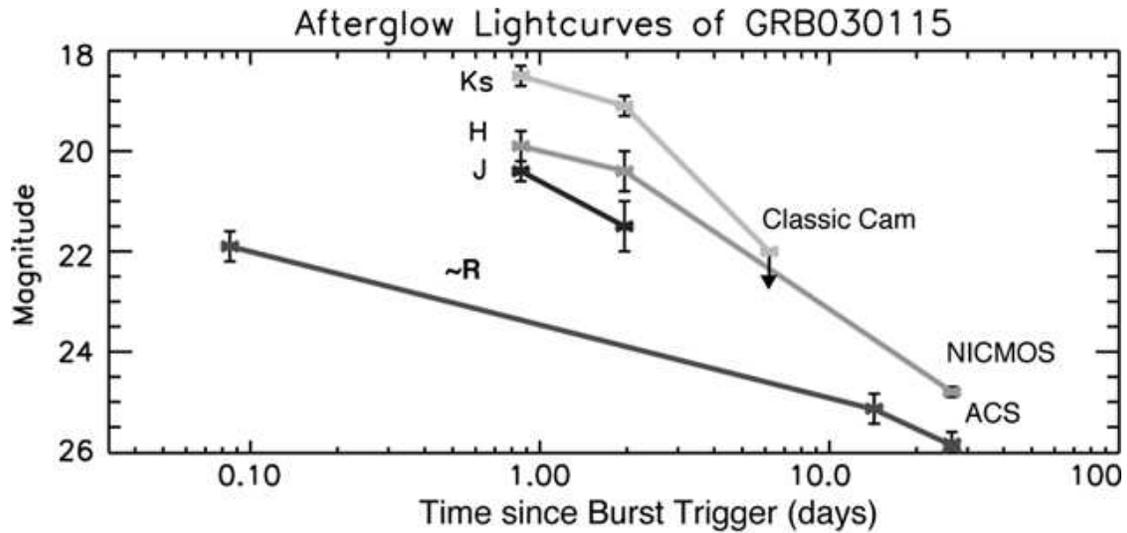}
  \caption{An updated lightcurve for GRB 030115, plotted using data from GCNs referenced below and including the new points from Classic Cam and HST.\label{fig3}}
\end{figure}


\bibliographystyle{aipproc}

\end{document}